# Synthetic Reflections on Resource Extraction

Sai Krishna [0009-0009-4034-8195], Vinaya Kumar [0000-0003-0106-3030],

Marc Böhlen [0000-0002-0463-0988]

University at Buffalo, Buffalo, NY 14260, USA

**Abstract.** This paper describes how AI models can be augmented and adapted to interpret landscapes. We present the technical framework of a Sentinel-2 satellite asset interpretation pipeline that combines statistical operations, human judgment, and generative AI models to produce succinct commentaries on industrial mining sites across the planet. To this end we introduce a novel bespoke landscape descriptor, the Urban Dwelling and Mining Index, and discuss how this metric can improve the performance of a multimodal language model in assessing the spatial distribution of mining operations.

## 1 Introduction

This planetary computing inquiry sets out to nudge older, yet powerful AI systems to gain an understanding of the impact of industrial level mining operations on landscapes across Earth.

Typically, extractivism, the economic model focused on removing large volumes of resources for financial gain, has a bad connotation. And rightly so. Extractivism is destructive, often irreversible and generates harm for the landscape and all creatures within it. At the same time, extractivism, in a fundamental way, as metabolic survival, is the very reason human beings have been able to exist and thrive on every continent on the planet. When we breathe, we extract oxygen from air that some other oxygen-dependent creature cannot have even when we exhale carbon dioxide that plants in turn use. Agriculture, a fundamental turning point in human civilization that created the first food surplus, specialization of labor and permanent settlements, is an extractive process at its core. The building of cities, the creation of machines, transportation networks and automation at large, was premised on massive extractivism and mining operations [10].

The point of this diversion is not to use a reductio ad absurdum argument to defend pathologic extractivism but rather to understand it as a path, regrettable in hindsight, that human beings have taken. In addition to investing in new futures that do not rely on extractivism, it makes sense to live up to the history we have created, and to take responsibility for our extractivist past.

One good use we can put the history of extractivism to, is to share it with others. More potently, in this inquiry, is the intent to share this history with industrialized AI systems.

Typically, AI is set in opposition to humankind, developing superintelligence [23, 25], devouring resources, while human beings are dispensable and perhaps an impediment to progress. By acknowledging the significance of extractivism and sharing this history with AI systems, we acknowledge a common denominator and articulate a concern for the future with AI systems dependent on the very concepts of extractivism created long before the start of AI systems.

This project sets out to translate that idea into a tangible artifact by documenting the scale of visible extractive practices such that AI can ingest the results. More specifically, the project maps out this shared extractive history by tallying hundreds of mining sites across the planet, interpreting satellite imagery from low Earth orbit of these sites with the assistance of AI models themselves, and then handing over the picture book to AI systems, with instructions on how to read the book.

This paper is informed of a longer inquiry into the fallout of planetary computing [3] and parallels a second study [4] in which we describe the project from the vantage point of the history of technology obsolescence. Here, we focus on the description of the methods devised to create the system, how we craft a pipeline from satellite imagery to image interpretation, and then integrate context sharing with other AI models with Agentic Retrieval Augmented Generation. Before we address our approach to AI-supported landscape interpretation, we will discuss our data sources.

## 2 Observation of Compromised Landscapes from Orbit

Earth Observation (EO) Geospatial AI is the synthesis of remote sensing and generative AI models. The field has received renewed attention with the recent contribution of DeepMind's AlphaEarth Foundations [12]. Unlike traditional methods that require massive datasets for each new task, AlphaEarth utilized self-supervised learning to reason with sparse data and to deliver embeddings of 64-dimensions that describe landscape features as a vector space.

Our project is invested in the possibility of using an AI model to interpret satellite imagery that can be understood by people, and as well as translated into vector space for machines. Moreover, we are interested in a computationally parsimonious approach that uses few resources and takes advantage of the churn of AI production that creates new models at rapid pace, leaving the no longer new but not old models on the sideline. The EU Sentinel-2 observation system fits this project philosophy.

The EU Sentinel-2 [15] orbiters are a unique set of low Earth orbit satellite assets. The Sentinel-2 project is part of the European Union's Copernicus Program dedicated to EO, designed to provide high-resolution, multispectral imagery of the Earth's land surfaces. Sentinel-2 consists of two identical satellites, Sentinel-2A and Sentinel-2B, and each satellite carries a multispectral instrument that captures data in 13 spectral bands, ranging from visible and near-infrared to short-wave infrared wavelengths, at spatial resolutions of 10–60 meters.

The widespread adoption of Sentinel-2 data in environmental monitoring and resource management is rooted in its ability to combine high revisit frequency with multi-spectral analysis at useful spatial analysis resolutions. In the realm of agriculture, these data are indispensable for precision farming; by leveraging specialized Red-Edge bands that are acutely sensitive to chlorophyll, the imagery allows for the calculation of advanced indices to monitor crop health, estimate yields, and detect early signs of plant stress. This level of detail extends into forestry and biodiversity assessment, where the 10-meter spatial resolution enables the mapping of land-cover changes, forest fragmentation, and habitat loss in near-real-time. Furthermore, the mission's capacity for frequent revisits—approximately every five days—provides a temporal continuity that is vital for tracking rapid environmental shifts, such as the spread of wildfires or the progression of harmful algal blooms in inland water bodies.

State of the art commercial EO systems such as Planet Labs [2], outclass Sentinel in terms of temporal and spatial resolution. Moreover, private sector providers offer on-demand tasking, enabling affluent customers to direct a satellite to capture specific locations at specific times, a service that public missions with fixed observation plans cannot provide. However, the Sentinel satellites remain a benchmark for dependable EO due to its high spectral richness, wide swath coverage and - most importantly for this project - its free and open data policy. This is important not only to mitigate financial resource constraints of NGOs, public institutions, and independent researchers, it also addresses the project's investment in technology reuse and resource parsimony. Finally, Sentinel-2 data policies support the transferability of the project and the potential for others to expand and continue the research.

The tradeoff between freely available data and spatial resolution is not one that negatively impacts what this project sets out to deliver. The temporal resolution limitations are inconsequential as mining activities are typically long term, industrial scale landscape alterations that take years and decades to complete. Additionally, because industrial mining sites are expansive, with large mines reaching multiple square kilometers in size, they can readily be observed from low-Earth orbit and the Sentinel-2 spatial limitations of 10m/pixel in the visible range, are not critical. In short, the impact of mining on the surface of Earth is significant enough to be readily observed with aging public sector low-Earth orbit remote sensing assets.

To create our shared history of extractivism, we built a database on mining sites from across the planet. At the time of this writing, the dataset contains about 200 distinct locations. We rely on geographic coordinates collected from an open-source collection of mining sites [31] and generate from these coordinates a bounding box of 100 square km around the center. We then select a time and date suitable for our project. That selection is informed of the conditions in the atmosphere - low cloud coverage - and on the ground. For example, in the northern hemisphere we constrain the search on snow free months in the spring and summer and set the inverse priorities for the southern hemisphere. We inform that constraint by the knowledge horizon of the specific AI models we are using, in the case of Llama-4, December 2024.

We then use the OpenEO framework [9] to collect Sentinel-2 assets that meet that requirement list. From that result, we preprocess each entry for data quality and then download the pruned list of candidates. Defined areas are not always covered in a single swath of a satellite pass, and orbital swath limits can lead to significant gaps at the edges of images. Image statistics can capture these conditions reliably. This pruned list is then further filtered based on image quality criteria including contrast, sharpness, and Shannon entropy, an approach routinely used in automated image quality assessment [8]. We average those scores while considering outliers. These steps, augmented by an automated image enhancement operation to increase contrast and color saturation, enable us to maximize the potential of the Sentinel-2 assets. However, the automation is imperfect, and each image is checked by a human to ensure that our mathematical operations produce an aesthetically acceptable result.

Because our project seeks to read the landscape and interpret the impact of surface mining on the environment, we make use of arithmetic operations on the rich Sentinel-2 data and select specific arithmetic operations on select bands. Our Sentinel-2 preparation pipeline creates visible images (RGB) as well as indices relevant to the reading of surface-level mining operations. We generate from the Sentinel-2 multi-band assets the Normalized Built-Up Difference Index (NDBI), the Normalized Difference Vegetation Index (NDVI), and the Ferrous Metal Index (FMI).

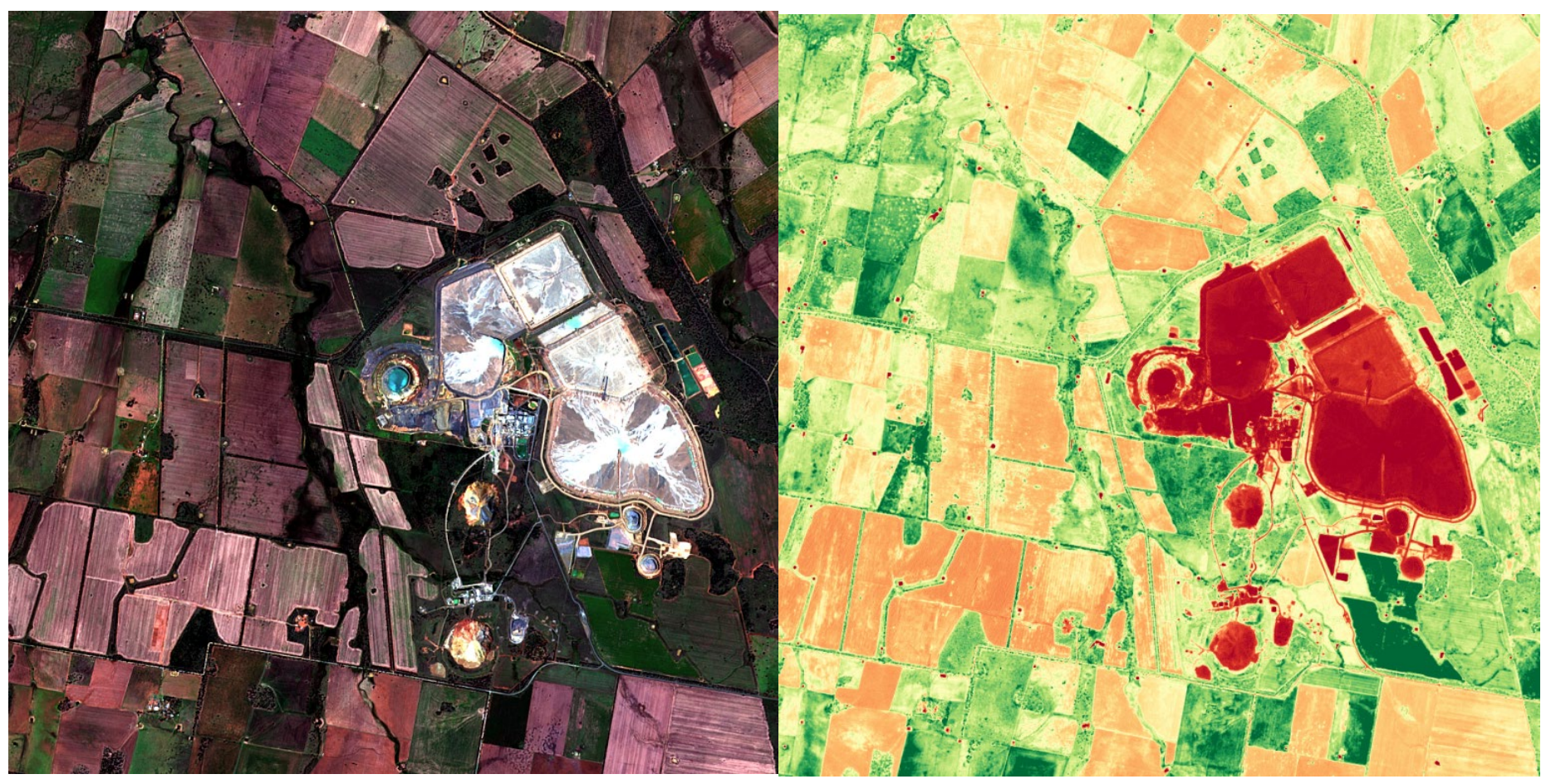

**Fig. 1.** Endeavour22, Northparkes Mine Project, Goonumbla, Kennedy Co., New South Wales, Australia. Left: Sentinel-2 RGB from 2024-12-27. Right: Corresponding NDVI interpretation. Red are areas with low vegetation scores.

The NDVI index is typically deployed to monitor vegetation. However, it is also quite effective at detecting large mining areas. Because surface mines tend to remove all vegetation in the active mining locations, the NDVI measures with uncanny precision the current outlines of an active mining operation as Fig. 1 demonstrates. However, NDVI cannot distinguish mining sites from urban development as both land use categories entail the removal of large vegetation swaths. The NDBI index, on the other hand, is dedicated specifically to the detection of urban built-up areas, but in turn it often conflates urban areas with mining sites.

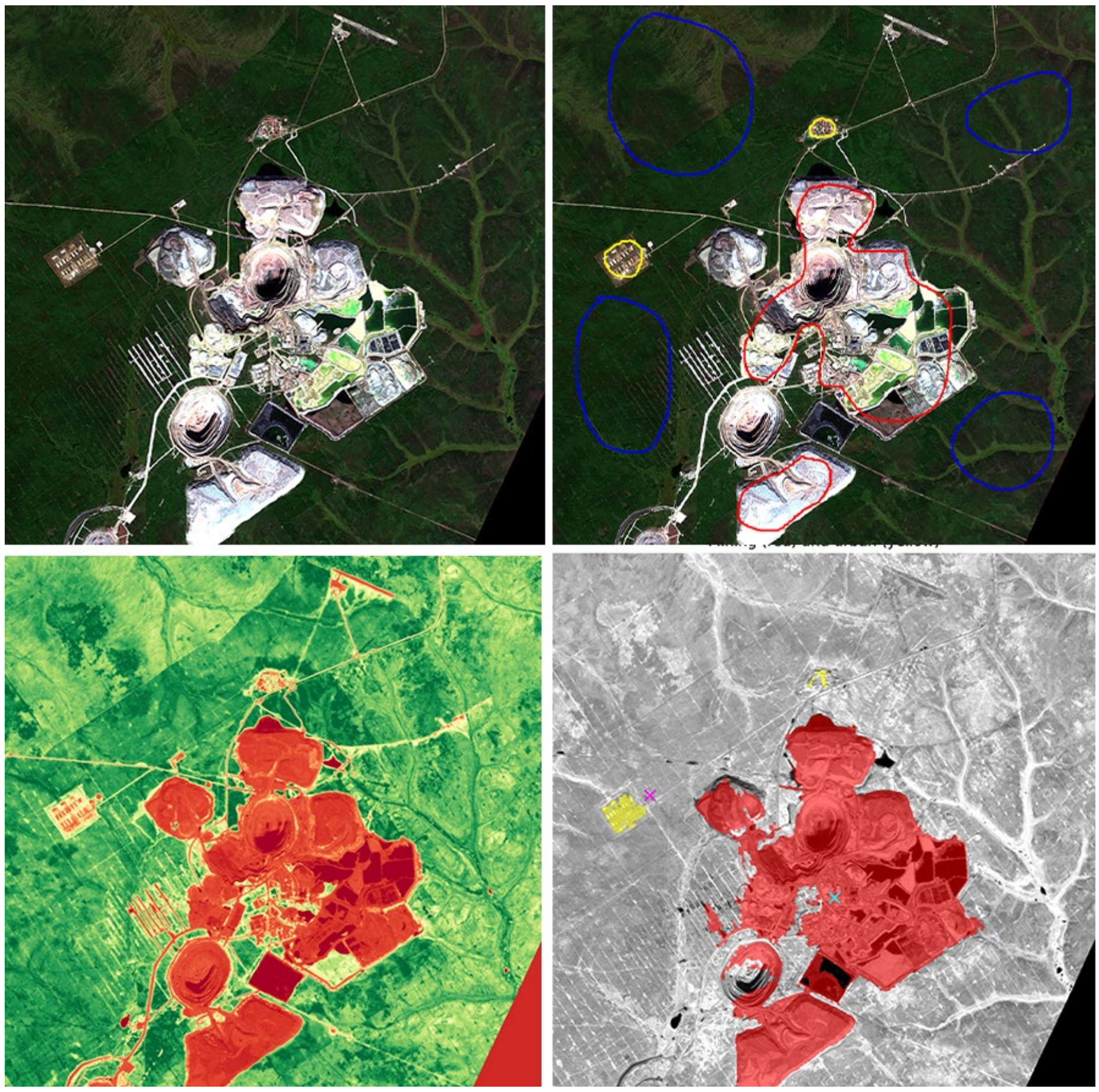

**Fig. 2.** Nyurbinsky open-pit mine, Sredne-Markhinsky region, Sakha, Russia. Sentinel-2 asset from 2024-08-17: Top row: RGB, Manual annotations, including negative samples in blue. Bottom row: NDVI and Urban Dwelling and Mining indices. The latter clearly identifies two housing clusters, one of which is small, and the extensive mining areas.

As we describe below, our approach requires a multi modal language model to interpret multi-spectral Sentinel-2 satellite data. Given that some models can ingest only RGB imagery, and the pertinent information on urban settlements and mining areas is not detectable in the three visible bands of the satellite image, we make use of the other bands, such as near infrared and short-wave infrared, to map salient information to an input space the language model can ingest.

For that reason, we created a bespoke index, the *Urban Dwelling and Mining Index* (UDM): a custom binary index for built-up areas and mining sites. This index relies on human help, annotations in the form of informal, but well-placed scribbles that denote the urban areas, the mining sites as well as landscape features that belong to neither category. This latter step is critical in enhancing the discriminating ability of this binary centroid classifier. The negative samples operate as veto to prevent false positives from urban-industrial landscapes with similar features. We build these components from 10 bands[1] of a Sentinel-2 geotiff asset to build spectral-texture features and check for edge cases such as mining pit interiors and shadows. Then we train centroid based binary classifiers using positive and negative samples. Minimum and maximum pixel areas allow us to adjust for particular urban settings, allowing for even small settlements to be considered. Morphology and NDVI gating criteria are added to reduce false positives. Fig.2 shows the process and result of the custom UDM index on Russia's Nyurbinsky Open Pit in the Sredne-Markhinsky region, Nakyn, Sakha, one of the largest and most productive diamond mines in the country. Even the small settlement in the middle of the image is properly detected while roads, as intended, are omitted. And the black swath at the image edge is properly recognized as irrelevant.

UDM is far from perfect and a formal evaluation of its capacity to address the distinction between the vast array of urban development configurations, formal and informal, remains to be investigated fully. Yet for our needs, the current version of the index proved helpful and undeniably superior to the default NDBI index typically used to map urban settlements. We will return to the UDM and its utility in our discussion at the end of the paper.

## 3 Multimodal AI Models

Large Language Models (LLMs) that defined the popularization of chatbots have foundational limitations regarding their ability to ingest and process information. Multimodal Large Language Models (MLLMs) extend traditional language-only models by processing and reasoning over multiple types of data—such as text, images, audio, video, and sensor signals—within a single unified framework [35].

[1] (B02, B03, B04, B05, B06, B07, B08, B8A, B11, B12). See GitHub repository for details.

Technically, MLLMs combine large language model backbones with modality-specific encoders such as vision encoders for images or audio encoders for speech. These components are aligned through joint training or fine-tuning so that information from different modalities can be mapped into an abstract, shared representation space. This allows the model to relate what it perceives through one modality to another modality: to relate imagery and audio to linguistic concepts.

The MLLM Kosmos-2 [30] serves as a foundational grounding model that pioneered the ability to link text descriptions to specific image regions via bounding boxes, effectively treating spatial coordinates as location tokens. Building on this legacy of integration, Meta's Llama-4 [26] suite utilizes a native multimodal, Mixture-of-Experts architecture to provide high-efficiency performance with massive context windows of up to 10 million tokens. Meanwhile, Anthropic's Claude [1] and OpenAI's GPT-5 [29] series represent the current frontier, merging rapid response capabilities with deep, chain-of-thought reasoning. However, GPT-5 is a very large model [5], and the performance of GPT-5 is energy intensive, consuming four to five times as much power as Llama-4 400B, aka Maverick in benchmark studies [22].

While MLLMs have the capacity to interpret imagery, models differ broadly in their ability to extract salient information, and the possibility of hallucinations or worse, AI Slop [20], is a constant distraction. While we experimented with the three different models, Kosmos-2, Llama-4 and GPT-5, we selected Llama-4 Maverick for its versatility and its ability to respond to our elaborate system prompt and multi-shot prompt collection through API calls to model repositories [28].

However, Llama-4 has a significant limitation, namely the fact that it can ingest only 3-band imagery. Frontier models such as GPT-5 can directly ingest high dimensional multi-spectral imagery. To address this limitation, we introduce several preprocessing steps to reduce the high multi-spectral imagery to subsets that contain pertinent spatial information and meet the input requirements of the Llama model. The next paragraphs describe the main features of the system prompt.

### 3.1 Nudging Model Behaviour with System Prompts

As our project seeks to read the landscape and interpret the impact of surface mining on the environment, we complement the above-mentioned band operations with instructions in our system prompt to the MLLM to facilitate the interpretation of the image processes. The band operations described above require context for interpretation, and our system prompt includes that important information.

While input prompts to MLLMs define the specific query, system prompts direct the AI system on how to generate its response. Core requirements for system prompts apply to most large language models, and some approaches are particular to individual models. Since our target here is the Llama series of MLLMs, we largely follow the recommendations outlined in related research [6].

Our system prompt includes details on the image indices created through the arithmetic operations as well as the on the custom designed binary classifier for the UDM index. We define the core requirements as a focus on the distribution of urban settlements, the location of mining areas, urban development, urban sprawl, and slums. Parallel to the first focus we add an environmental condition focus, to wit water conditions, disturbed land surfaces, vegetation health, and environmental stress relying on what can be detected in the supplied imagery. We augment those requirements with a description of landscape specific signatures, a caption template, and an output format. Importantly, we define a series of constraints to nudge the text generation in our desired direction and limit fabrications, To tune the linguistic quality of the result, we instruct the MLLM to refrain from vague quantifiers, reduce the use of exaggeration, avoid repetitive phrases, and vary diction where possible. While our final version is fine-tuned to one specific application, the approach we deploy and the categories we use to achieve the goals extend to other contexts.

In addition to the imagery, query, and system prompt, we provide our candidate MLLM with a context statement, metadata, that we describe in the next section.

### 3.2 Purposeful Context

Our system should be able to cogently interpret satellite imagery of mining sites. However, interpretation includes more than merely diligently reading the landscape, it includes considering conditions related to the direct observations, we want our system to include, where possible the history of the site.

To that end, we create for each site a collection of metadata with information that the imagery as supplied does not directly include. The core human-produced contribution to MLLM input sequence is the collection of metadata assets, one for each mining site. For each mining site we collect information on the geology of the area, the history of the mining site and controversies on the mining operation that have been recorded. We make use of a variety of sources for this three-part metadata collection. Our sources are site specific and can contain information from Wikipedia, mindat.org [31], mining company corporate documents [17], reports from the Global Energy Monitor [18], reports from think tanks, and regional environment protection agencies [27]. We also include, where available, scholarly research that addresses land use rights as well as documented impacts on water, air, and soil quality. We then compile these diverse sources into 250-word text segments, where each segment contains information on the mine’s history, the area’s geology, and contentious issues. Each metadata set is combined with the other inputs described above for direct ingestion into the MLLM pipeline.

To be clear, publicly available information on mining sites is not equally distributed across the planet, and countries that restrict public opinions also tend to have fewer sources of information on mining activities available. Our system prompt attempts to control the MLLM in such cases to prevent unwarranted inventiveness. The GitHub repository [32] contains in addition to our code base the metadata collection deployed in this project.

# 4 AI to Evaluate AI

Assessing the quality of a machine-produced artifact is a challenge for AI systems as well as for human beings. This project relies on a combination of human judgement and AI-based analytics. As mentioned above, each Sentinel-2 asset is evaluated analytically and then subject to a review by a human expert in visual culture. Likewise, the text output produced by the MLLM is evaluated analytically and by the human research team. While the human image evaluation is rapid due to evolutionary efficiency of human visual processing, evaluating text is far more tedious. Moreover, our initial experiments showed that statistical metrics such as BLEU and BERT Score [7] were inadequate to detect the kind of textual nuance we seek in our texts. To address the text evaluation bottle neck, we use AI to evaluate AI. Specifically, we integrate an older LLM into our system and instruct it to gauge whether the MLLM's output text is adequate and follow existing approaches [19] to that end. Specifically, we make use of LLM Gemini-Flash [16], recently replaced by the newer Gemini 3 [13] models.

Our instructions to the LLM include a multi-shot instruction set with multiple examples. Moreover, we include a list of five requirements, namely: a) focus on environmental conditions, b) use of context-specific terminology, c) observation of patterns, d) adherence to constraints, and e) conciseness. Each of these conditions are defined in the system prompt supplied to the LLM. Each candidate caption is evaluated on a five-point scale for its adherence to these categories and deemed acceptable when a set threshold is achieved for the average across all five conditions. Additionally, our decision rule adds a quality gating condition, namely a minimum per-dimension score requirement. Together, these rules prevent a small set of strong performances to override failures in other dimensions. For example, with this approach the evaluator LLM prevents the MLLM-generated caption text with a high score in conciseness but a low score in (the more important category of) environmental framing from being accepted.

To reduce ambiguity and improve machine-readability, we require the evaluator LLM to return scores as a structured JSON object with fixed keys and numeric values. We implement this using a structured-output interface (JSON MIME type and a schema that specifies required fields and types).

Despite using schema-constrained structured outputs, we observed evaluation responses that were not valid JSON or did not conform to the expected structure (e.g., additional prose around the object, markdown wrapping, missing keys, or minor formatting corruption such as trailing commas or inconsistent quoting). Our initial setup asked the evaluation module to score all five rubric dimensions in a single call. Pilot experiments revealed a systematic artifact: rubric items at the beginning and end of the list more often received stable, complete scores, while intermediate dimensions were more likely to be shallow, missing, or inconsistent. To mitigate this, we moved to per-criterion evaluation, scoring each rubric dimension in a separate judge call. While this increased the number of API calls, it substantially improved completeness, reliability of rubric scoring and reduced partial-completion and omission artifacts. Integrating all these procedures into our LLM as a critic resulted in a robust filter for poorly formulated image captions that we can deploy at scale.

To be clear, the eloquence of the machine-produced texts leaves something to be desired. Despite the harsh imagery supplied by our Sentinel-2 system, the tone of voice of our output is cool and analytical. This is the price we pay for preferencing accuracy with this AI model. Perhaps this stolid tone fits the gravity of the topic. Yet even with this elaborate evaluation scheme seeking to optimize accurate visual image interpretation, the LLM at times failed to recognize features a human can easily detect. In one case, for example, the caption text reported on the distribution of mining sites in the image, confusing sparse cloud cover with small mines, and the automated check did not find fault with the description.

## 5 Results

As described above, we experimented with three different MLLM systems in this project. The simplest of the models, Kosmos-2, failed to differentiate more nuanced landscape features. The descriptions also tended towards vague and generic formulations that could not meet our project requirements. GPT-5, the most complex of the three systems and by far the largest model, delivered excellent results. Moreover, GPT-5 (Pro) supports the ingestion of native geotiff datasets, allowing us to add an entire Sentinel-2 spectral cube containing 13 dimensions as input. Importantly, GPT-5 can process the attended image metadata, allowing it to access the precise geolocation references contained within the geotiff. No other information beyond the spectral cube was made available to the GPT-5 model.

Our experiments with Llama-4 required substantially more effort and pre-processing labor as described above. As opposed to GPT-5, Llama-4 is limited to 3-band image inputs. The addition of the NDVI landscape interpretation allowed Llama-4 to make more robust observations on the distribution of vegetation, and to be more precise on the boundaries of vegetation-poor areas, a feature most active mining sites share. Interestingly the additional capacity to evaluate vegetation did not negatively impact the model's ability to reason about the extent of mining sites.

Here is an example of how GPT-5 described the Kolwezi Mine in DR Congo:

> Sentinel-2 (11 October 2024) shows Kolwezi, DR Congo. Large, stepped excavations and pale dumps dominate the frame. Light-toned haul roads thread between cut slopes, stockpiles and angular pads. Several bright turquoise basins and channels lie within the industrial footprint. Surrounding surfaces are largely bare, with vegetation confined to scattered patches; no large natural water body is evident. Visible environmental impacts include land-cover conversion, fragmentation, exposed soils and altered surface-water pathways around engineered basins and roads. The scene documents extensive disturbed ground across a mining district, with sharp boundaries between worked areas and remaining cover emphasized by shadowed benches and embankments.

We experimented with hyperparameters to tune our MLLM model of choice towards our desired performance regime. Changes to the hyperparameter temperature which controls randomness of the model's output were not particularly effective in producing richer language and increased unwanted fabrications. We made use of a frequency penalty parameter together with enforcement of word selection to reduce token-level repetition, resulting in fewer repeated phrases. Additionally, we achieved a significant improvement in the fluency and coherence as well as consistency, relevance and even found sparks of nuance with a multi-shot approach. We included in our system prompt multiple (5 to 10) sample question-answer pairs to serve as templates for the text generation. These samples were crafted to meet our requirements in content and tone and designed in accordance with the system prompt's list of requirements, constraints and preferences. The refinement of the system prompt across almost a dozen iterations, as well as the addition of the above-mentioned landscape indices, nudged Llama-4 almost to the level of GPT-5-like response at a fraction of the compute cost. Processing time and compute requirements factored into our evaluation scheme as well. GPT-5 required over 10 minutes, and in some cases over 20 minutes to produce its artifacts while our reduced-complexity Llama-4 approach delivered solid results in under 60 seconds. Given the pricing for GPT-5, the choice of which model to settle on resolved itself quickly.

The overview below shows samples from our experiments in augmenting the input to the MLLM Llama-4 with additional landscape indices. The top section shows the RGB image-only result, and the bottom section the RGB+NDVI+ UDM result. Both versions received the same context data. We observed that more detailed observations on the distribution of urban dwellings are produced when the UDM index is included (italicized below).

Garzweiler Open Pit, North Rhine-Westphalia, Germany.

*[RGB only]*
This Sentinel-2 image from October 23, 2024, shows the Garzweiler open-cast lignite mine in North Rhine-Westphalia, Germany. The landscape is dominated by the large open pit, with extensive excavation and removal of lignite, a type of brown coal. The surrounding area is a mix of agricultural land and fragmented forests. The mine's footprint is broad, with visible benches and spoil heaps indicating active extraction. Environmental indicators include significant land degradation, soil erosion, and potential groundwater contamination due to the large-scale excavation. The nearby agricultural lands and settlements are at risk from dust and chemical pollution. The expansion of the mine has led to the destruction of villages and infrastructure, highlighting the social and environmental impacts of lignite mining.

*[RGB+NDVI+UDM]*
The October 23, 2024 Sentinel-2 image captures the Garzweiler open-cast lignite mine in North Rhine-Westphalia, Germany, operated by RWE Power. The mine, one of Europe's largest lignite surface mines, is surrounded by a patchwork of agricultural fields and fragmented forests. The extensive open-pit operation, covering about 48 square kilometers, shows significant land disturbance with exposed soil and bare rock. The image data indicates severe vegetation loss and stress around the mine. *The Urban Dwelling and Mining Index highlights the mining footprint, showing a large area dedicated to extraction*. The environmental impact includes potential soil erosion, water pollution, and habitat loss. *Nearby communities may face risks from dust, noise, and potential water contamination*. As Germany plans to end lignite mining by 2030, the mine's expansion raises concerns about its environmental and social legacy.

Nyurbinsky Open Pit, Sakha, Russia

*[RGB only]*
The Sentinel-2 image from August 17, 2024, depicts the Nyurbinsky Open Pit diamond mine in Russia's Sakha Republic. The scene is dominated by a large-scale open-pit operation, with multiple pits and associated infrastructure, including tailings ponds and haul roads, amidst a largely forested landscape. The mining footprint is extensive, with several large pits and waste rock dumps visible, indicating significant land disturbance.
The environmental indicators suggest severe land degradation, with exposed soil and bare rock prevalent around the mining areas. The presence of tailings ponds with potentially contaminated water poses a risk to nearby water bodies and groundwater. Vegetation stress is evident in areas adjacent to the mining operations, likely due to soil erosion and potential chemical contamination.
The risk of environmental degradation is heightened by the mine's proximity to the Vilyuy River basin, a sensitive water system. The extraction of diamonds at this site contributes to the observed environmental impacts, including land disturbance and potential water pollution. The ongoing mining activities pose a risk of further environmental degradation, emphasizing the need for careful management to balance industrial output with environmental protection.

*[RGB+NDVI+UDM]*
The Nyurbinsky Open Pit diamond mine, captured on August 17, 2024, by Sentinel-2, is situated in a remote, forested region of Russia's Sakha Republic. The mine's infrastructure, including multiple open pits and processing facilities, is visible amidst the surrounding green landscape. Image data indicates significant land disturbance and vegetation loss around the mining areas. The mine's footprint is extensive, with large open pits and associated infrastructure, showing a broad footprint of mining activity. *A localized urban area is present near the mining operations.*
The environmental indicators suggest severe vegetation stress and loss around the mine, with potential soil erosion and habitat disruption. The risk of water pollution and ecosystem degradation is possible due to the mine's proximity to local water systems linked to the Vilyuy River basin. The extraction of diamonds at this site may lead to long-term environmental consequences for the surrounding ecosystem and potentially impact local communities.

Loolekop Pipe Palabora Mine, Limpopo, South Africa

| |
|---|
| *[RGB only]*<br>This Sentinel-2 image from November 18, 2024, shows the Loolekop Pipe Palabora Mine in South Africa, a unique carbonatite-hosted copper mine. The image reveals extensive open-pit mining and associated infrastructure, with large areas of disturbed land and bare soil indicating significant environmental degradation. The mine, which also yields phosphate, iron, and vermiculite, has a substantial footprint with multiple pits and tailings areas. A large tailings pond is visible, with potentially contaminated water posing a risk to nearby water bodies. The surrounding landscape shows sparse vegetation, likely due to mining activities and soil erosion. The environmental impact is evident in the stressed vegetation and altered land surfaces around the mine. The proximity of the mine to Kruger National Park highlights potential ecological risks and tensions between industrial activities and environmental conservation. |
| *[RGB+NDVI+UDM]*<br>The Loolekop Pipe Palabora Mine in South Africa, captured on November 18, 2024, by Sentinel-2, shows extensive mining activity with multiple open pits and tailings ponds. The mine, known for extracting copper, phosphate, iron, and vermiculite, has caused substantial land disturbance and vegetation loss. Image data indicates severe vegetation stress around the mine and tailings ponds, with large areas of bare substrate and disturbed terrain. *A small urban area is visible to the northeast*. Environmental concerns include soil erosion, water pollution from tailings, and habitat fragmentation. The proximity of the mine to nearby settlements and Kruger National Park may lead to water contamination for people living in the vicinity. The extensive footprint of the mine and associated infrastructure underscores the need for careful management to mitigate these impacts. |

## 6 AI for People

We devised a resource-constrained interface to simultaneously view the distribution of mining sites across the planet included in our dataset as well as the generated landscape interpretations of individual mining sites as shown in Fig. 3. Each mining site on the gray and solemnly rotating Earth is marked as an orange disk. Clicking on any of these disks launches a visualization of the respective mining site, together with the interpretation of the satellite data.

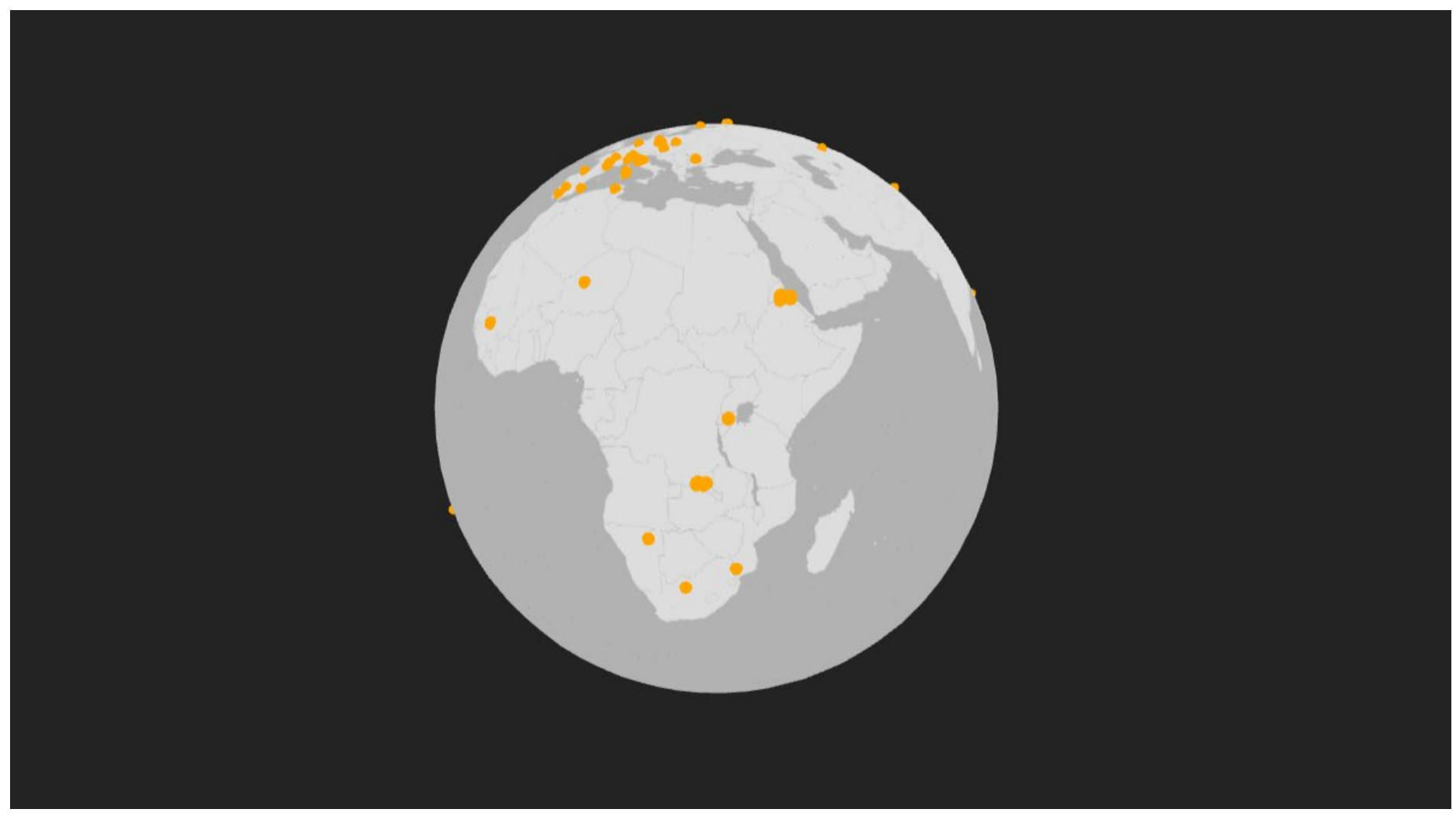

**Fig. 3.** The rotating Earth as interface (video: https://tinyurl.com/ScrapyardAI ).

## 7 AI for AI

While our mining site observations and landscape interpretation collection is far from complete, we are considering how it might be shared not only with people, but with other AI systems. It might be possible to include the materials as training data for future frontier models, but that option is not available to us, as we lack the resources to operate on frontier models.

Instead of retraining large models at prohibitive cost, we can offer our data collection as an external source to other models and help them make use of the assets.

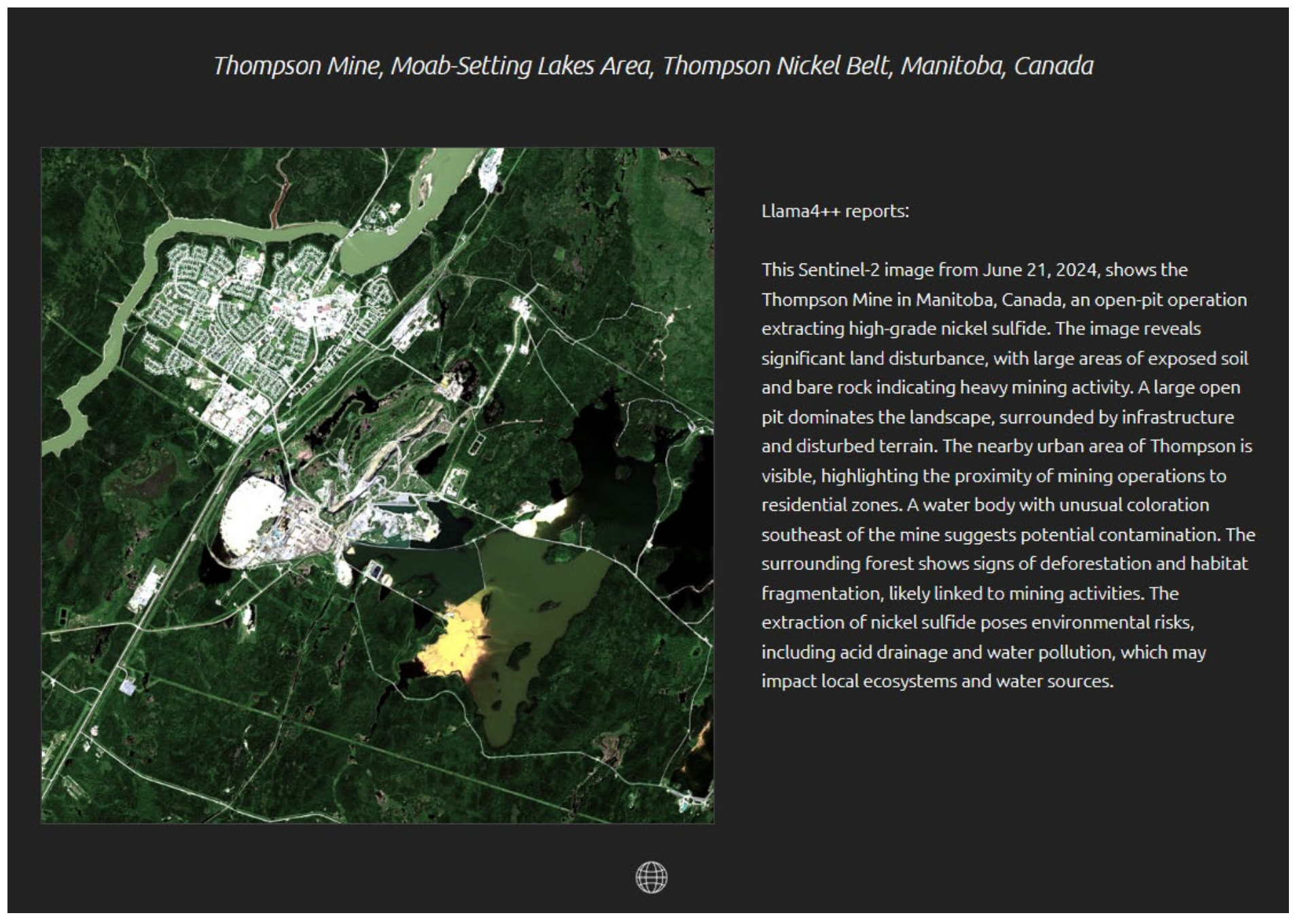

**Fig. 4.** A multimedia description of the generated observations on the Thompson Mine in Manitoba, Canada.

### 7.1 Retrieval-Augmented Generation

A Retrieval-Augmented Generation (RAG) architecture transforms raw text into a queryable knowledge base and is widely used in scientific literature analysis, and decision-support systems. Rather than relying solely on parameters learned during training, a RAG system retrieves relevant documents or passages from an external knowledge source—such as a document collection—based on a query [24]. These retrieved texts are supplied as context to an LLM, which produces responses grounded in the supplied evidence. This approach reduces hallucinations and improves factual reliability, particularly for domain-specific knowledge. However, RAG systems can struggle to aggregate large datasets effectively [21] and fail to find common threads across individual data entries.

In this work, we apply a RAG framework to synchronize unstructured satellite observations with a structured vector-space index, bridging the gap between the static knowledge of an LLM and the dynamic nature of satellite-caption data. To preserve semantic integrity, the pipeline employs a recursive character splitter operating in token space, using 150-token chunks with a 30-token overlap to maintain contextual continuity. Prior to embedding, chunks are enriched with structured metadata—such as mine name, country, and geolocation—prepended directly to the text to improve discoverability. The augmented chunks are then encoded by a sentence-transformer model into high-dimensional embeddings, which are stored in a vector database alongside their original text and metadata.

At query time, the system follows a two-stage retrieval–generation process. The query is embedded using the same transformer model to ensure alignment, and a top-k similarity search retrieves the most relevant records. Retrieved chunks provide both evidentiary text and associated metadata (e.g., mine identifiers, location, and filenames), which are concatenated into a unified context block for the generative model. To minimize hallucination, a closed-domain system prompt constrains the model to generate responses strictly from this context and to reference specific metadata, anchoring all claims to verifiable sources. This process allows us to connect our mining site observations with LLMs that have not been previously exposed to the knowledge base, extending the reach of our world view, as it were.

### 7.2 Agentic Retrieval-Augmented Generation

Our implementation extends the standard RAG to an *Agentic RAG* [11]. Agentic RAGs differ from standard RAG though their ability to perform multi-step reasoning and self-evaluate its response. Here we give the system additional long-form context and nudge the system to perform cross-site reasoning. For each query, the agent retrieves from both stores and consolidates the most relevant results into a single evidence set. Because long-form documents can obscure critical passages or lose section-level context, we index them using a three-layer hierarchy as shown in Fig. 6. A document map captures metadata and section headers for routing, embedded section summaries support retrieval at the level of arguments and claims, and atomic evidence chunks retain pointers to their parent sections. At query time, the agent executes a coarse-to-fine retrieval cascade, selecting relevant sections before extracting precise evidence. If the retrieved context is insufficient, the agent refines the query and repeats retrieval before producing a closed-domain, evidence-grounded response. Additionally, we added a fuzzy matcher to associate country names with adjective forms, a case the RAG pipeline managed unreliably.

To evaluate the Agentic RAG approach, we added long form contextual data; a book-length publication on the impacts of mining on Indigenous communities in remote Australia [33]. The text included below shows how the Agentic RAG nudges the LLM DeepSeek [14] to describe mining sites in Australia with both specific details *and* observations informed of pertinent sections of the book mentioned above.

> Mining operations in Australia have caused profound and lasting environmental degradation across multiple sites. At the Ranger Uranium Mine in the Kakadu region, operations have led to over 120 regulatory infringements, including contaminated water releases from tailings dams into the Magela Creek system, directly impacting Indigenous water sources. The Mount Pleasant Open Cut Mine in New South Wales, a massive six-square-kilometer black coal operation, has caused extensive habitat destruction and ecosystem disruption. Historical mines like Mary Kathleen in Queensland, which extracted uranium and rare earth oxides, have left a legacy of radioactive waste and severe land degradation, with visible bare soil and challenges for revegetation decades later. The extraction of sulphide-rich ores at sites like the historic Elliots No. 1 Open Cut contributes to acid mine drainage, leading to long-term soil erosion and chemical contamination of waterways. These impacts are not isolated; they represent a pattern where large-scale land disturbance, water pollution, and habitat loss are direct consequences of mining activity.

----------------------------------------------------------------------------------------------------------

Caption Sources:
    ElliotsNo1OpenCut, Endeavour22, CentralNorthOpenPit
Document Sources:
    Scambary_MyCountryMyMine_2013_239p.pdf > Page 178
    Scambary_MyCountryMyMine_2013_239p.pdf > Page 178
    Scambary_MyCountryMyMine_2013_239p.pdf > Page 3

**Fig 5.** Response (accepted answer) of the LLM DeepSeek-Chat, combined with our Agentic RAG, to the query: "How do mining operations in Australia impact the environment? Elaborate on specific examples." The log information below the dash line specifies where the Agentic RAG found pertinent information used in the response.

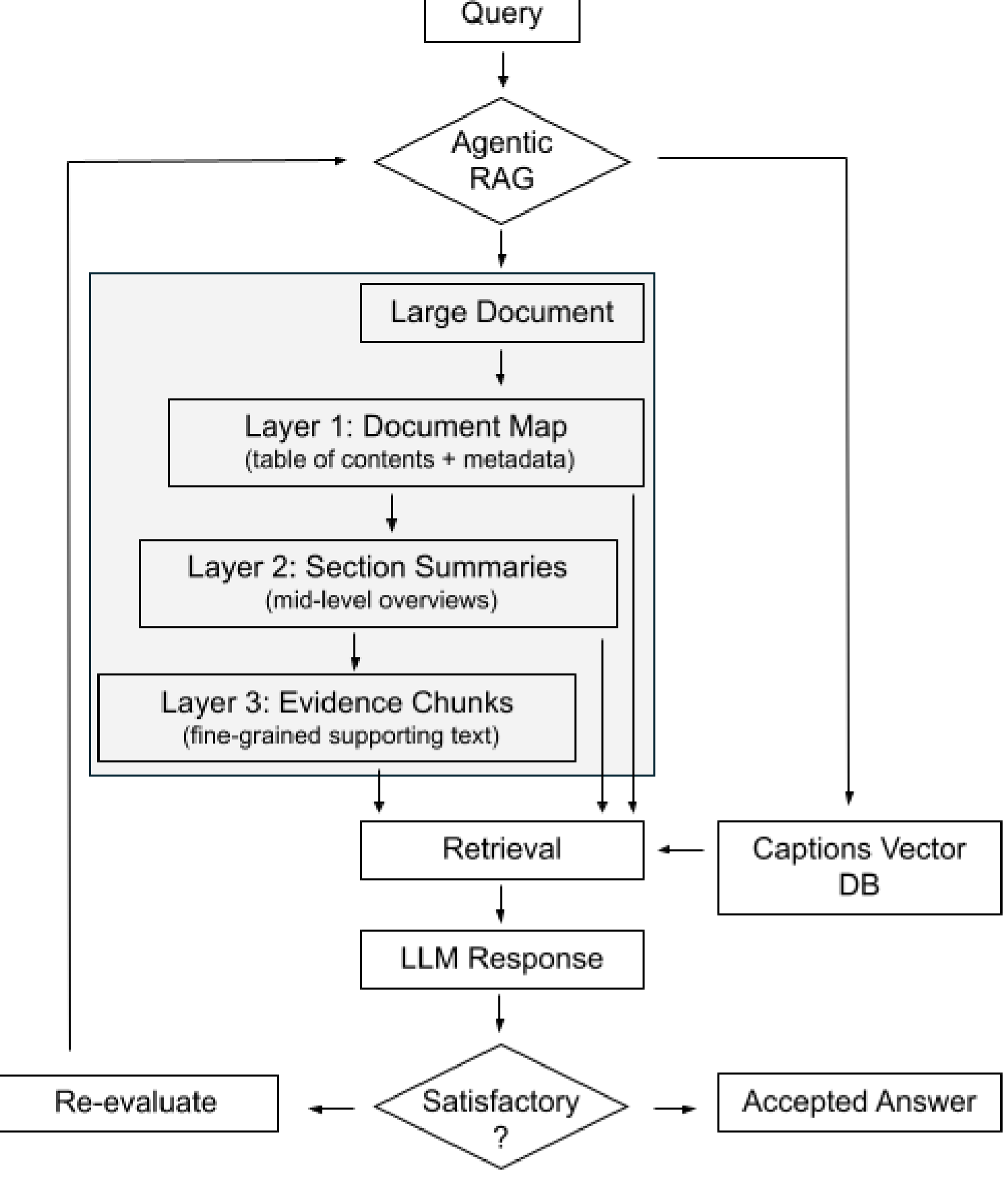

**Fig 6.** Diagram of the Agentic RAG and its connection to an external LLM.

# 8 Discussion

Increasingly powerful AI models are typically used to perform tasks people do and in many cases replace human labor, effectively engaging in long-term confrontation between AI and people. These same systems, however, can document an uncomfortable past and present that is shared between people and AI systems. Considering the history of planetary resource extraction as well as our own limited resources leads to new sites of inquiry, namely, the increasingly large collection of no longer new yet powerful AI models.

Operating from this AI-induced cultural condition, we set out to build a system that could observe, interpret and comment on large-scale mining operations. While we succeeded in this respect, we are far from general-purpose landscape interpretation, that requires much more nuance. That extended project, perhaps through an integration of AlphaEarth [12], remains an open task.

Human-in-the-loop paradigms [34] have been suggested as frameworks to ensure that human agency in AI artifacts can remain intact. This project suggests one path to enable the human-in-the-loop in the current GenAI landscape: mix and match human agency and AI model components to a joint mixture of experts. While our Agentic RAG experiment shows that in principle we can nudge one particular external LLM to incorporate our landscape interpretation knowledge base and produce results that address both specific details in the datasets as well as observations across details, we do not know how successful the approach fares with other models, and how it might scale in the future, as frontier models continue to increase in size and complexity.

Finally, the research project articulates an opportunity to nudge AI model output, even modestly, to articulate planetary-scale environmental concerns and economize on computational resources at the same time.